\documentclass[prc,aps,reprint,superscriptaddress,showpacs]{revtex4-1}
\usepackage{amsmath}
\usepackage{graphicx}
\usepackage{url}
\usepackage{dcolumn}
\usepackage{bm}
\usepackage{units}

\begin{document}
\title{Cross-section measurement of the $^{130}$Ba(p,$\gamma$)$^{131}$La reaction for $\gamma$-process nucleosynthesis}

\author{L.~Netterdon}
\email{lnetterdon@ikp.uni-koeln.de}
\affiliation{Institute for Nuclear Physics, University of Cologne, Z\"ulpicher Stra\ss e 77, D-50937 Cologne, Germany}
\author{A.~Endres}
\affiliation{Institute for Nuclear Physics, University of Cologne, Z\"ulpicher Stra\ss e 77, D-50937 Cologne, Germany}
\affiliation{Institut f\"ur Angewandte Physik, Goethe Universit\"at Frankfurt am Main, D-60438 Frankfurt am Main, Germany}
\author{G.~G.~Kiss}
\affiliation{Institute for Nuclear Research (MTA ATOMKI), H-4001 Debrecen, POB.51., Hungary}
\author{J.~Mayer}
\affiliation{Institute for Nuclear Physics, University of Cologne, Z\"ulpicher Stra\ss e 77, D-50937 Cologne, Germany}
\author{T.~Rauscher}
\affiliation{Centre for Astrophysics Research, School of Physics, Astronomy and Mathematics, University of Hetfordshire, Hat AL10 9AB, United Kingdom}
\affiliation{Departement f\"ur Physik und Astronomie, Universit\"at Basel, Klingelbergstrasse 82, Basel CH-4056, Switzerland}
\author{P.~Scholz}
\affiliation{Institute for Nuclear Physics, University of Cologne, Z\"ulpicher Stra\ss e 77, D-50937 Cologne, Germany}
\author{K.~Sonnabend}
\affiliation{Institut f\"ur Angewandte Physik, Goethe Universit\"at Frankfurt am Main, D-60438 Frankfurt am Main, Germany}
\author{Zs.~T\"or\"ok}
\affiliation{Institute for Nuclear Research (MTA ATOMKI), H-4001 Debrecen, POB.51., Hungary}
\author{A.~Zilges}
\affiliation{Institute for Nuclear Physics, University of Cologne, Z\"ulpicher Stra\ss e 77, D-50937 Cologne, Germany}

\begin{abstract}
%\begin{description}
\setlength{\parindent}{0pt}
\textbf{Background:} Deviations between experimental data of charged-particle induced reactions and calculations within the statistical model are frequently found. An extended data base is needed to address the uncertainties regarding the nuclear-physics input parameters in order to understand the nucleosynthesis of the neutron-deficient $p$ nuclei.

\textbf{Purpose:} A measurement of total cross-section values of the $^{130}$Ba(p,$\gamma$)$^{131}$La reaction at low proton energies allows a stringent test of statistical model predictions with different proton+nucleus optical model potentials. Since no experimental data are available for proton-capture reactions in this mass region around A~$\approx$~130, this measurement can be an important input to test the global applicability of proton+nucleus optical model potentials.

\textbf{Method:} The total reaction cross-section values were measured by means of the activation method. After the irradiation with protons, the reaction yield was determined by use of $\gamma$-ray spectroscopy using two clover-type high-purity germanium detectors. In total, cross-section values for eight different proton energies could be determined in the energy range between \unit[3.6]{MeV} $\leq E_p \leq$ \unit[5.0]{MeV}, thus, inside the astrophysically relevant energy region.

\textbf{Results:} The measured cross-section values were compared to Hauser-Feshbach calculations using the statistical model codes \textsc{TALYS} and \textsc{SMARAGD} with different proton+nucleus optical model potentials. With the semi-microscopic JLM proton+nucleus optical model potential used in the \textsc{SMARAGD} code, the absolute cross-section values are reproduced well, but the energy dependence is too steep at the lowest energies. The best description is given by a \textsc{TALYS} calculation using the semi-microscopic Bauge proton+nucleus optical model potential using a constant renormalization factor. 

\textbf{Conclusions:} The statistical model calculation using the Bauge semi-microscopic proton+nucleus optical model potential deviates by a constant factor of 2.1 from the experimental data. Using this model, an experimentally supported stellar reaction rate for proton capture on the $p$ nucleus $^{130}$Ba was calculated. At astrophysical temperatures, an increase in the stellar reaction rate of \unit[68]{\%} compared to rates obtained from the widely used \textsc{NON-SMOKER} code is found. This measurement extends the scarce experimental data base for charged-particle induced reactions, which can be helpful to derive a more globally applicable proton+nucleus optical model potential.
%\end{description}
\end{abstract}


\maketitle

\section{Introduction}
\label{sec:introduction}
The synthesis of the heavy elements beyond the iron-peak region remains a partly open question in nuclear astrophysics. About \unit[99]{\%} of the heavy elements are produced during the $s$- and $r$-processes \cite{Kaeppeler11,Arnould07}. However, a small fraction of neutron-deficient nuclei are bypassed by these neutron-capture processes. They are denoted as $p$ nuclei \cite{Arnould03,Rauscher13}. These about 35 proton-rich nuclei in the mass region between Se and Hg are believed to be produced by a variety of different processes, usually summarized as the $p$ process. Among others, astrophysical processes producing the $p$ nuclei are the $\gamma$ process in type II supernovae \cite{Rayet90,Rauscher02} and type Ia supernovae \cite{Travaglio11}, the $rp$ process during thermonuclear burning on a neutron-star surface \cite{Schatz98}, or the $\nu p$ process in neutrino-driven winds of type II supernovae \cite{Froehlich06}. Lighter $p$ nuclei are also efficiently produced in type Ia supernoave \cite{Travaglio11, Kusakabe11}.

Up to current knowledge, the majority of the $p$ nuclei are produced by photodisintegration reactions during the $\gamma$ process within O/Ne burning layers of core-collapse supernovae. When the shock-front passes the O/Ne layer, temperatures of \unit[2]{GK}~$\leq T \leq$~\unit[3.5]{GK} are reached, allowing the partial photodisintegration of pre-existing seed nuclei. The $\gamma$-process starts with sequences of ($\gamma$,n) reactions. At some point, the ($\gamma$,n) reactions will start to compete with ($\gamma$,p) and ($\gamma$,$\alpha$) reactions as well as $\beta$ decays, leading to deflections in the $\gamma$-process path.
The reaction rates in the $\gamma$-process reaction network, which includes thousands of reactions on mainly unstable nuclei, are calculated within the scope of the Hauser-Feshbach (HF) statistical model \cite{Feshbach52}. In order to obtain reliable model predictions, it is important to put the nuclear-physics input parameters entering these calculations on a firm basis. These nuclear-physics input parameters include nuclear level densities and $\gamma$-ray strength functions which determine the $\gamma$ width. Moreover, the particle+nucleus optical-model potentials (OMP) are needed to describe the particle widths for protons, neutrons, and $\alpha$-particles. These parameters can, to some extent, be experimentally tested by laboratory experiments.

Experimental data at astrophysically relevant energies are generally scarce. Besides the activation technique, which has been widely used for the determination of total cross sections before \cite{Yalcin09,Kiss11,Dillmann11,Sauerwein11,Halasz12,Netterdon13}, also the in-beam technique with high-purity germanium (HPGe) detectors \cite{Galanopoulos03,Sauerwein12,Harissopulos13,Netterdon14} and the 4$\pi$-summing method \cite{Tsagari04,Spyrou07,Simon13} is available. In order to systematically check the validity of the HF calculations, experimental data over a wide mass range is highly desired. This has motivated measuring the cross-section values of radiative proton-capture on the $p$ nucleus $^{130}$Ba. Especially in the mass region around $A~\approx~130$ no experimental data at low interaction energies are available for proton-capture reactions. 

\begin{figure}[tb]
\centering
\includegraphics[width=\columnwidth]{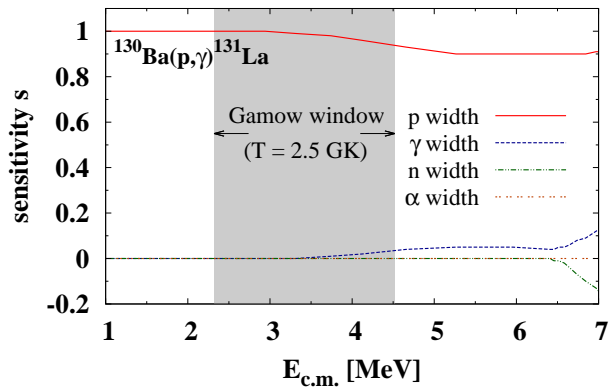}
\caption{(Color online) Sensitivity s of the $^{130}$Ba(p,$\gamma$)$^{131}$La reaction cross-sections, when the proton (p), neutron (n), $\alpha$, and $\gamma$ widths are varied by a factor of two, as a function of center-of-mass energies. The shaded area denotes the Gamow window for a temperature of \unit[2.5]{GK}. Within the measured energy region the cross section is dominantly sensitive to the proton width.}
\label{fig:sensitivity}
\end{figure}

Figure~\ref{fig:sensitivity} shows the sensitivity of the $^{130}$Ba(p,$\gamma$) laboratory cross section for the charged-particle widths, neutron width, and $\gamma$ width as a function of center-of-mass energy. The sensitivity $s$ denotes the relative change of the cross section when one width is varied \cite{Rauscher12}:
\begin{equation}
s = \frac{\frac{\sigma'}{\sigma}-1}{f-1} \, ,
\end{equation}
where $\frac{\sigma'}{\sigma}$ is the relative variation of the laboratory cross section and $f=\frac{\Gamma'}{\Gamma}$ denotes the factor, by which the respective width is varied. Within the investigated energy region, the $^{130}$Ba(p,$\gamma$) reaction cross section is dominantly sensitive to the proton width and, thus, the proton+nucleus OMP. Therefore, the cross section is only affected by changes of the proton-OMP. Hence, the present measurement is well-suited to test HF model predictions using different proton+nucleus OMPs.

\section{Experiment}

The reaction $^{130}$Ba(p,$\gamma$)$^{131}$La was investigated by means of the activation method. Cross-section values were measured at center-of-mass energies between \unit[$E_{\mathrm{c.m.}}\,=\,3.57$]{MeV} and \unit[$E_{\mathrm{c.m.}}\,=\,4.96$]{MeV}. The maximum of the reaction rate integrand is located at an energy of \unit[3.3]{MeV} for a temperature of \unit[2.5]{GK}. This is expected to be the relevant temperature for the nucleosynthesis of $^{130}$Ba during the $\gamma$ process \cite{Rauscher13}. At this temperature, the astrophysical Gamow window ranges from \unit[$E_{\mathrm{c.m.}}\,=\,2.32$]{MeV} to \unit[$E_{\mathrm{c.m.}}\,=\,4.52$]{MeV} \cite{Rauscher10}. Thus, the cross-section values were measured at an energy range overlapping with the Gamow window. 

The irradiations as well as the $\gamma$-ray spectroscopy of the activated targets were carried out at the Institute for Nuclear Physics in Cologne, Germany. After activation periods of \unit[3]{h} to \unit[5]{h}, the $\gamma$ activity of the activated reaction products was detected using two clover-type HPGe detectors.

\subsection{Target properties}

\begin{figure}[tb]
\centering
\includegraphics[width=\columnwidth]{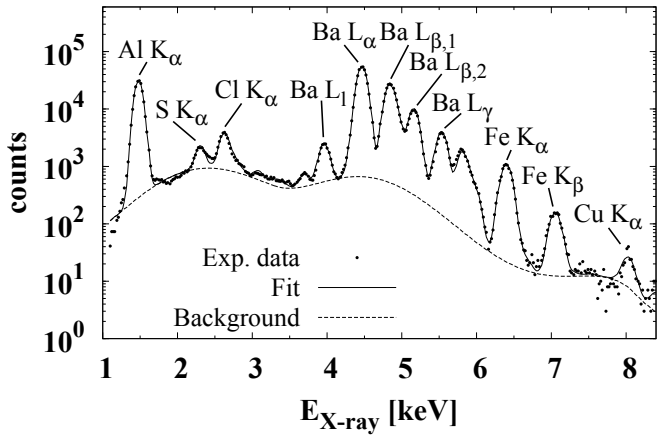}
\caption{X-ray spectrum obtained from a PIXE measurement for one of the Ba targets. One can clearly identify the characteristic Al K$_\alpha$ transition stemming from the backing material as well as various X-ray transitions of Ba. Moreover, some lighter contaminants like Fe and Cu can be seen. The dashed line depicts the simulated continuous bremsstrahlung background.}
\label{fig:pixe}
\end{figure}

Three targets were prepared by evaporating  BaCO$_3$ onto \unit[2]{$\mu$m}-thin Al foils. The Ba content of the carbonate powder was enriched to \unit[(11.8~$\pm$~0.2)]{\%} in $^{130}$Ba. The same targets have already been used for an $\alpha$-induced measurement on $^{130}$Ba, see Refs.~\cite{Halasz12,Gyurky11} for details of the target production process. The target thicknesses were remeasured by means of proton-induced X-ray emission (PIXE) at ATOMKI, Debrecen, prior to this experiment. The uncertainty of this PIXE measurement amounts to \unit[$\pm$5]{\%}. Figure~\ref{fig:pixe} shows an X-ray spectrum obtained during the PIXE measurement for one of the used Ba targets. Within the given uncertainties of the earlier results, namely \unit[$\pm$6]{\%} (RBS), \unit[$\pm$8]{\%} ($\alpha$-energy loss), and \unit[$\pm$7]{\%} (weighing), no deviation was found. The final amount of target nuclei was obtained from the weighted average of these results. Target thicknesses were measured to be \unit[$\left(554.4 \pm 17.6\right)$]{$\frac{\mu \mathrm{g}}{\mathrm{cm}^2}$}, \unit[$\left(698.3 \pm 21.9\right)$]{$\frac{\mu \mathrm{g}}{\mathrm{cm}^2}$}, and \unit[$\left(989.9 \pm 30.7\right)$]{$\frac{\mu \mathrm{g}}{\mathrm{cm}^2}$}, respectively.
Taking into account the enrichment in the BaCO$_3$ powder, this leads to areal particle densities of $^{130}$Ba nuclei of \unit[$\left(37.9 \pm 1.2\right) \times 10^{15}$]{cm$^{-2}$}, \unit[$\left(47.8 \pm 1.5\right) \times 10^{15}$]{cm$^{-2}$}, and \unit[$\left(67.7 \pm 2.1\right) \times 10^{15}$]{cm$^{-2}$}, respectively.
Irradiated targets could be reused for later irradiations due to the short half-life of the reaction product, thus, the preparation of three targets was sufficient for this experiment. After the target preparation process, a protective Au layer was evaporated on top of each target to avoid deterioration of the target material. These Au layers have thicknesses of \unit[$22$]{$\frac{\mu \mathrm{g}}{\mathrm{cm}^2}$} and \unit[$67$]{$\frac{\mu \mathrm{g}}{\mathrm{cm}^2}$}, respectively. The target thicknesses were also measured during the irradiations by means of Rutherford Backscattering Spectrometry (RBS). Within the given uncertainties, no deterioration of target material was observed.

\subsection{Experimental setup}

\begin{figure}[tb]
\centering
\includegraphics[width=\columnwidth]{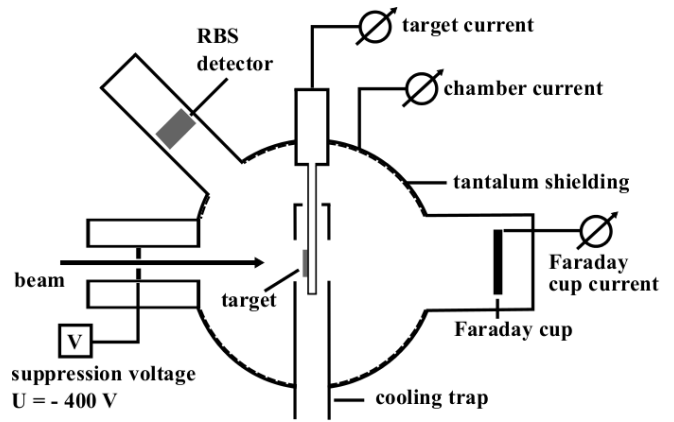}
\caption{Sketch of the target chamber used for nuclear astrophysics experiments. The suppression voltage of \unit[$U~=~-400$]{V} at the entrance is used to suppress secondary electrons in order to guarantee a reliable charge collection. The target itself is surrounded by a cooling trap at liquid nitrogen temperature to minimize residual gas deposits on the target. A silicon detector is also provided to measure the target thickness during the experiment by means of RBS.}
\label{fig:chamber}
\end{figure}

The proton beam was delivered by the \unit[10]{MV} FN tandem ion accelerator at the Institute for Nuclear Physics in Cologne. Figure~\ref{fig:chamber} shows the target chamber designed for nuclear astrophysics experiments which was used for this experiment. The deposited charge is measured at three different positions. It is measured at the target, at the chamber itself to measure released secondary electrons as well as scattered beam particles, and at the Faraday cup behind the target. The beam was stopped in a thick Au backing directly behind the target. Thus, no charge was measured at the Faraday cup. The deposited charge is measured using current integrators with an uncertainty of about \unit[4]{\%}. A negatively charged aperture with a voltage of \unit[$U\,=~\,-400$]{V} prevents secondary electrons from leaving the target chamber.
The target is surrounded by a cooling trap which is cooled down by LN$_2$ to reduce residual gas deposits on the target material. Moreover, the target chamber houses a silicon detector, which is used for RBS measurements throughout the experiment. By this, the target stability and thickness can be monitored during the irradiations.

The energy of the proton beam impinging on the target was determined by scanning the \unit[$E_p\,=\,3674.4$]{keV} resonance of the $^{27}$Al(p,$\gamma$) reaction \cite{Brenneisen95}. From the sharp rising edge of the obtained resonance yield curve, a spread in proton energy of \unit[$\pm4$]{keV} as well as a constant offset of \unit[29]{keV} with respect to the literature value of the resonance was found. Details of this procedure can be found in Ref.~\cite{Netterdon14}. This proton-energy spread and offset is taken into account for the determination of the proton energy and energy loss in the target material.

\subsection{Irradiation and $\gamma$ spectroscopy}
\label{sec:counting}
The beam current ranged from \unit[100]{nA} to \unit[180]{nA} for the different proton energies. For the highest beam energy, the beam current was limited to \unit[25]{nA} due to technical issues. The Q value of the $^{130}$Ba(p,$\gamma$)$^{131}$La reaction amounts to \unit[$\left(3796.5 \pm 28.1\right)$]{keV} \cite{QCalc}. The irradiations lasted from \unit[3]{h} to \unit[5]{h}. Since the unstable reaction product has a rather short half-life of $T_{1/2} = $~\unit[$\left(59 \pm 2\right)$]{min} \cite{NNDC}, the irradiated targets could be reused for later irradiations. No remaining radioactivity was present after a waiting time of about \unit[7]{h}.

\begin{figure}[tb]
\centering
\includegraphics[width=\columnwidth]{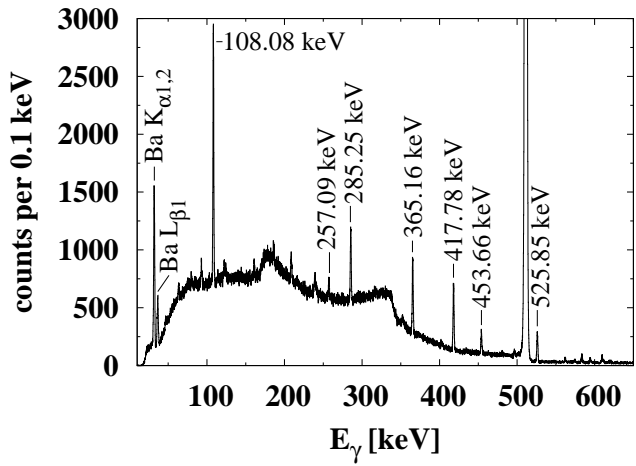}
\caption{Relevant part of a summed $\gamma$-ray spectrum measured with two HPGe clover detectors. This target was irradiated with \unit[4.8]{MeV} protons. The dominant transitions used for data analysis are highlighted. Moreover, characteristic X-rays of Ba are clearly visible at the low-energy end of the spectrum.  This spectrum was recorded for about \unit[3.5]{h}. The remaining $\gamma$-ray transitions stem from naturally occurring background. The $\gamma$-ray transition with \unit[$E_\gamma$~=~108.08]{keV} was not used for data analysis, see text for details.}
\label{fig:spectrum}
\end{figure}

The spectroscopy of the subsequent electron-capture decay of $^{131}$La was performed off-beam using two clover-type HPGe detectors, which are arranged in a head-to-head geometry. Each provides a relative efficiency of \unit[100]{\%} at a $\gamma$-ray energy of \unit[$E_\gamma\,=\,1.33$]{MeV} compared to a standard \unit[7.62]{cm}~$\times$~\unit[7.62]{cm} NaI detector. The detectors are surrounded by a \unit[10]{cm}-thick lead wall and a \unit[3]{mm}-thick copper sheet to suppress the natural background and X-rays stemming from the lead. In order to obtain a preferably high full-energy peak efficiency, the counting distance from the target to detector end cap was only \unit[1.3]{cm}. Owing to the rather short half-life of the reaction product, the $\gamma$-ray spectroscopy typically started approximately \unit[20]{min} after the end of the irradiation. Figure~\ref{fig:spectrum} shows a $\gamma$-ray spectrum recorded for about \unit[3.5]{h} using the clover-setup from a target irradiated with \unit[4.8]{MeV} protons. The six most probable $\gamma$-ray transitions subsequent to the electron-capture decay of $^{131}$La were used for the data analysis. Their properties are given in Table~\ref{tab:transitions}. The weaker $\gamma$-ray transitions as visible in Fig.~\ref{fig:spectrum} stem from naturally occurring radioactivity.

After the electron-capture decay of $^{131}$La, an isomeric state with a half-life of $T_{1/2}$\:=\:\unit[(14.6~$\pm$~0.2)]{min} at an excitation energy of $E_x\,=\,$\unit[187.50]{keV} is populated. This state decays by emitting $\gamma$ rays with energies $E_\gamma\,=~$\unit[79.9]{keV} and $E_\gamma\,=\,$\unit[108.08]{keV} in a cascade. The population probability of this isomeric state after the electron-capture decay was found to be less than \unit[1]{\%} \cite{Horen63,Harmatz72}. However, the observed $E_\gamma\,=\,$\unit[108.08]{keV} transition, also indicated in Fig.~\ref{fig:spectrum}, is then partly fed by the decay of the higher-lying isomeric state. Therefore, this $\gamma$-ray transition was not used for data analysis.

\begin{table}[tb]
\caption{Decay data of the reaction product $^{131}$La. Only $\gamma$-ray energies and absolute intensities of transitions used for the data analysis are listed. The decay parameters were adopted from Ref.~\cite{NNDC}.}
\label{tab:transitions}
\begin{ruledtabular}
\setlength{\tabcolsep}{1cm}
\begin{tabular}{rr}

\multicolumn{1}{c}{$E_\gamma$ [keV]}	&	\multicolumn{1}{c}{$I_\gamma$} \\

\colrule
 & \\
257.087 $\pm$ 0.009	&	3.428 $\pm$ 0.068 \\
285.246	$\pm$ 0.007 &	12.400 $\pm$ 0.028 \\
365.162 $\pm$ 0.008 &	16.925 $\pm$ 0.033\\
417.783 $\pm$ 0.015 &	17.950 $\pm$ 0.040 \\
453.659 $\pm$ 0.015 &	5.875 $\pm$ 0.125 \\
525.851 $\pm$ 0.016 &	8.725 $\pm$ 0.175 \\

\end{tabular}
\end{ruledtabular}
\end{table}

\subsection{Detector efficiencies}

\begin{figure}[tb]
\centering
\includegraphics[width=\columnwidth]{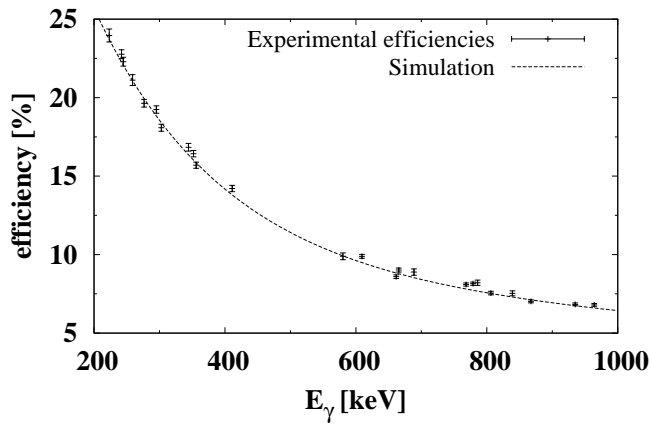}
\caption{Summed experimental full-energy peak efficiencies of the two HPGe clover detectors for a target-to-detector distance of \unit[1.3]{cm}, $i.e.$, the used counting distance. To account for coincidence-summing effects, the efficiencies were measured at a distance of \unit[10]{cm} where coincidence summing is negligible, and subsequently scaled to the counting distance of \unit[1.3]{cm} using a $^{137}$Cs source. The experimental data are compared to a \textsc{Geant4} simulation without summing effects and a very good agreement is found. See text for further details.}
\label{fig:efficiency}
\end{figure}

The absolute full-energy peak efficiency of the clover-type HPGe detectors must be known for the determination of the reaction cross-section. Four different calibrated radioactive sources, $^{133}$Ba, $^{137}$Cs, $^{152}$Eu, and $^{226}$Ra were used for this purpose. The low activity of the activated targets necessitated a short target-to-detector distance of \unit[1.3]{cm} to obtain a preferably high full-energy peak efficiency. At such a short distance, coincidence-summing effects can be significant, when $\gamma$-rays are emitted in a cascade, which would falsify the measured efficiencies. Thus, the absolute full-energy peak efficiencies of the clover setup were measured at a larger target-to-detector distance of \unit[10]{cm}. At this distance, coincidence-summing effects are negligible. Using a $^{137}$Cs source, where no cascading $\gamma$-ray transitions occur, a conversion factor was derived between the short and large target-to-detector distance. With the conversion factor obtained this way, the measured full-energy peak efficiencies were scaled to the counting distance, which are shown in Fig.~\ref{fig:efficiency}. Simulations using the \textsc{Geant4} \cite{Geant4,Allison06} toolkit confirmed that this conversion factor is energy-independent between \unit[200]{keV}~$\leq E_\gamma \leq$~\unit[2000]{keV}, $i.e.$, in the energy region of interest. Moreover,  \textsc{Geant4} simulations with single $\gamma$ rays were performed for the close geometry and good agreement was found for the comparison to the experimentally obtained efficiencies without coincidence summing effects.
Figure~\ref{fig:efficiency} shows the experimental full-energy peak efficiency at a distance of \unit[1.3]{cm} compared to the \textsc{Geant4} simulation. As a last step, the $\gamma$-ray cascades of the $^{131}$Ba $\gamma$ decay were implemented into the \textsc{Geant4} simulation to investigate possible coincidence summing effects. For the present case, the summing effects were found to be negligible. 

For data analysis, the full-energy peak efficiencies were obtained by fitting a function of the form 
\begin{equation}
f(E_\gamma) = a \cdot \exp(b \cdot E_\gamma) + c \cdot \exp(d \cdot E_\gamma)
\end{equation}
to the experimental data. The statistical uncertainty stemming from the fit is approximately \unit[2]{\%} for all $\gamma$-ray energies used for data analysis.

\section{Experimental results and discussion}

The effective center-of-mass energies were calculated by taking into account the proton energy loss in the protective Au layer and target material. The energy loss was calculated using the \textsc{Srim} code \cite{Srim12} and amounts to \unit[25]{keV} to \unit[57]{keV} depending on the proton energy and target thickness. The effective center-of-mass energy was determined by

\begin{equation}
E_{\mathrm{c.m.}} = E_p - \frac{\Delta E}{2} \, ,
\end{equation}
where $\Delta E$ denotes the energy loss in the target and $E_p$ the incident proton energy. This is appropriate, if the cross section changes only slightly over the target thickness. This is valid for the present case, since the cross-section uncertainties are larger than the changes of the cross-section prediction over the target thickness. The energy straggling inside the target material ranged from \unit[7]{keV} to \unit[11]{keV} and was added to the energy spread of the proton beam by means of Gaussian error propagation. 

The measured cross-section values are listed in Table~\ref{tab:results}. The results are given for each $\gamma$-ray transition, when applicable, as mentioned in Sec.~\ref{sec:counting}, as well as their weighted average. Of the uncertainty reported, \unit[5]{\%} stems from the detection efficiency, \unit[3]{\%} from the target thickness, \unit[4]{\%} from the charge collection, and \unit[2]{\%} to \unit[14]{\%} from statistical uncertainties. The statistical uncertainty for the cross-section value at \unit[$E_{\mathrm{c.m.}}$~=~4978]{keV} is significantly larger due to a lower beam current during the experiment resulting in a very low $\gamma$ activity.

\begin{table*}
\caption{Summary of the experimental cross-section values for each center-of-mass energy $E_{\mathrm{c.m.}}$. The respective areal particle densities $m$ of the irradiated targets are also listed for each energy. When available, the total cross section obtained using different $\gamma$-ray transitions are given, as well as the weighted average. The uncertainty is given by the variance of the weighted mean.}
\label{tab:results}
\begin{ruledtabular}
\begin{tabular}{cccccc|ccccc}
$E_{\mathrm{c.m.}}$ [keV] & $m$ [10$^{15}$ cm$^{-2}$] & $E_\gamma$ [keV]	&	$\sigma$ [mb]	& $\bar{\sigma}$ [mb] & & $E_{\mathrm{c.m.}}$ [keV] & $m$ [10$^{15}$ cm$^{-2}$]& $E_\gamma$ [keV]	&	$\sigma$ [mb]	& $\bar{\sigma}$ [mb] 	 \\
\colrule
 & & & & & & & \\
3573 $\pm$ 11& 47.8~$\pm$1.5	&	257.09   &	-	& 	0.25 $\pm$ 0.02		&&	4381 $\pm$ 8& 37.9~$\pm$1.2	&	257.09   &	1.47 $\pm$ 0.13			& 	1.47 $\pm$ 	0.11 	\\
			&	&	285.25	 &	0.26 $\pm$ 	0.02& 			&&				&	&	285.25	 &	1.46 $\pm$ 0.12			&			\\
			&	&	365.16   &	0.24 $\pm$	0.02& 			&&				&	&	365.16   &	1.50 $\pm$ 0.12			&			\\
			&	&	417.78 	 &	0.23 $\pm$	0.02& 			&&				&	&	417.78 	 &	1.51 $\pm$ 0.11	        &			\\
			&	&	453.66	 &	0.28 $\pm$	0.03& 			&&				&	&	453.66	 &	1.45 $\pm$ 0.12	 	    &			\\
			&	&	525.85   &	0.23 $\pm$	0.02& 			&&				&	&	525.85   &	1.45 $\pm$ 0.12			&			\\
			&	&			 &					&			&&				&	&			 &				&			\\
3785 $\pm$ 9& 37.9~$\pm$1.2	&	257.09   &	-	& 	0.35 $\pm$ 0.03		&&	4570 $\pm$ 12&67.7~$\pm$~2.1	&	257.09	 &	2.48 $\pm$ 0.19			&	2.44 $\pm$ 0.17		\\
			&	&	285.25	 &	0.37 $\pm$ 0.03	& 			&&				&	&	285.25	 &	2.37 $\pm$ 0.18			&			\\
			&	&	365.16   &	0.36 $\pm$ 0.03	& 			&&				&	&	365.16	 &	2.48 $\pm$ 0.19			&			\\
			&	&	417.78 	 &	0.32 $\pm$ 0.03	& 			&&				&	&	417.78	 &	2.36 $\pm$ 0.18			&			\\
			&	&	453.66	 &	-				& 			&&							&	&	453.66	 &	2.48 $\pm$ 0.19			&			\\
			&	&	525.85   &	0.35 $\pm$ 0.03	& 			&&				&	&	525.85	 &	2.47 $\pm$ 0.18			&			\\
			&	&			 &					&			&&				&	&			 &							\\
3972 $\pm$ 12& 67.7~$\pm$~2.1	&	257.09   &	0.73 $\pm$ 0.06	& 	0.74 $\pm$ 0.05	&&	4776 $\pm$ 9&47.8~$\pm$1.5	&	257.09   &		3.55 $\pm$ 0.27		& 	3.77 $\pm$ 0.28			\\
			&	&	285.25	 &	0.73 $\pm$ 0.06	& 			&&				&	&	285.25	 &		3.77 $\pm$ 0.29		&			\\
			&	&	365.16   &	0.80 $\pm$ 0.07	& 			&&				&	&	365.16   &		3.84 $\pm$ 0.30		&			\\
			&	&	417.78 	 &	0.72 $\pm$ 0.06	& 			&&				&	&	417.78 	 &		3.71 $\pm$ 0.29     &			\\
			&	&	453.66	 &	0.74 $\pm$ 0.06	& 			&&				&	&	453.66	 &	 	4.00 $\pm$ 0.31	    &			\\
			&	&	525.85   &	0.73 $\pm$ 0.06	& 			&&				&	&	525.85   &		3.72 $\pm$ 0.29		&			\\
			&	&			 &					&			&&				&	&			 &				&			\\				
4179 $\pm$ 9&47.8~$\pm$1.5	&	257.09   &	0.97 $\pm$ 0.07	& 	0.97 $\pm$ 0.07	&&	4978 $\pm$ 8& 37.9~$\pm$1.2	&	257.09   &	5.03 $\pm$ 0.80			& 	4.70 $\pm$ 0.68			\\
			&	&	285.25	 &	0.94 $\pm$ 0.08	& 			&&				&	&	285.25	 &		4.63 $\pm$ 0.73		&			\\
			&	&	365.16   &	0.96 $\pm$ 0.08	& 			&&				&	&	365.16   &		4.24 $\pm$ 0.66		&			\\
			&	&	417.78 	 &	0.96 $\pm$ 0.07	& 			&&				&	&	417.78 	 &		4.57 $\pm$ 0.72     &			\\
			&	&	453.66	 &	0.99 $\pm$ 0.07	& 			&&				&	&	453.66	 &	 	4.65 $\pm$ 0.73	    &			\\
			&	&	525.85   &	0.99 $\pm$ 0.07	& 			&&				&	&	525.85   &		4.98 $\pm$ 0.79		&			\\

\end{tabular}
\end{ruledtabular}
\end{table*}

\begin{figure}[tb]
\centering
\includegraphics[width=\columnwidth]{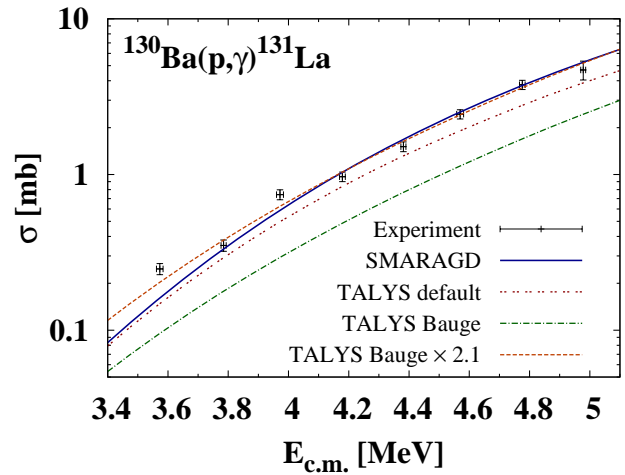}
\caption{(Color online) Experimental cross section of the $^{130}$Ba(p,$\gamma$)$^{131}$La reaction as a function of center-of-mass energy. The total cross sections are compared to theoretical predictions obtained from the \textsc{SMARAGD} code \cite{Smaragd}, using the proton+nucleus OMP from Ref.~\cite{Jeukenne77} with low-energy modifications from Ref.~\cite{Lejeune80}. A reasonable description of the experimental data is obtained ($\chi^2_{\mathrm{red}}~=~4.38$), but the energy dependence is too steep at the lowest energies. Furthermore, the experimental values are compared to calculations using the \textsc{TALYS} code \cite{Talys}, using the phenomenological proton+nucleus OMP from Ref.~\cite{Koning03} (\textsc{TALYS} default), as well as the semi-microscopic OMP from Ref.~\cite{Bauge01} (\textsc{TALYS} Bauge). Scaling the `\textsc{TALYS} Bauge' cross-section values by a factor of 2.1 yields on average a better description of the energy dependence at low energies ($\chi^2_{\mathrm{red}}~=~2.30$). Details about the used proton+nucleus OMPs can be found in the text.}
\label{fig:results}
\end{figure}

Figure~\ref{fig:results} shows the experimentally obtained cross sections in comparison with theoretical calculations using the statistical model codes \textsc{SMARAGD} \cite{Smaragd} and \textsc{TALYS 1.4} \cite{Talys}. As shown in Fig.~\ref{fig:sensitivity}, the laboratory cross section in the measured energy range is almost exclusively sensitive to the proton width and, thus, to the proton+nucleus optical model potential (OMP). In the following, the reduced $\chi^2$ values were calculated with 7 degrees of freedom, $i.e.$ the number of experimental data points minus 1. The potential from Ref.~\cite{Jeukenne77} with low-energy modifications from Ref.~\cite{Lejeune80} is used in the \textsc{SMARAGD} calculation. A reasonable agreement is found for this calculation ($\chi^2_{\mathrm{red}}~=~4.38$). However, the energy dependence is predicted to be too steep at the low-energy region. This would lead to an underestimation of the reaction rate, when extrapolating into the astrophysical energy region. 
In addition, a recently suggested modification of the imaginary part of the proton+nucleus OMP from Refs.~\cite{Jeukenne77,Lejeune80} was tested on the $^{130}$Ba(p,$\gamma$) reaction. This modification improved the reproduction of measurend proton-induced reaction cross sections for lighter nuclei \cite{Kiss07,Kiss08,Rauscher09}. However, for the present case this modified proton+nucleus OMP leads to very similar cross-section values as the unmodified version. 

Statistical model calculations were additionally performed using the code \textsc{TALYS}. Fig.~\ref{fig:results} shows a \textsc{TALYS} calculation using the semi-microscopic proton+nucleus OMP of Ref.~\cite{Bauge01} (`\textsc{TALYS} Bauge'). Please note, that this proton+nucleus OMP is usually denoted as `JLM' in \textsc{TALYS}. In this case, the energy dependence is reproduced very well, but on an absolute scale, the cross-section values are underestimated by about a factor 2 ($\chi^2_{\mathrm{red}}~=~55.13$). Scaling these theoretical cross-section values by a factor of 2.1 improves the $\chi^2_{\mathrm{red}}$ significantly and yields $\chi^2_{\mathrm{red}}~=~2.30$. In the present case, it is appropriate to scale the cross section instead of the widths entering the statistical-model calculation, since the cross section is only sensitive to the proton width, see Fig.~\ref{fig:sensitivity}. The scaled cross-section values obtained with the semi-microscopic proton+nucleus OMP of Ref.~\cite{Bauge01} are also shown in Fig.~\ref{fig:results} (`\textsc{TALYS} Bauge $\times$ 2.1').

Both semi-microscopic proton+nucleus OMPs have a common microscopic approach. They are based on Brueckner-Hartree-Fock approximation \cite{Brueckner64} with Reid's hard core nucleon-nucleon interaction \cite{Reid68} and adopting a local density approximation. However, beside the fact, that different theoretical nuclear densities are used, the optical-model parameters obtained by fitting to experimental data are different as well. Especially, in the case of Ref.~\cite{Bauge01}, the isovector component is renormalized by a factor of 1.5 compared to the proton+nucleus OMP from Ref.~\cite{Jeukenne77}. This renormalization was introduced in Ref.~\cite{Bauge01} to account for neutron and proton elastic scattering and (p,n) cross sections. However, in the present case for a radiative proton-capture reaction at sub-Coloumb energies, it strongly underestimates the experimental cross-section values, although the energy dependence is predicted correctly.\\
Moreover, \textsc{TALYS} calculations were performed using its default settings, where the phenomenological proton+nucleus OMP from Ref.~\cite{Koning03} is used. This calculation is denoted as `\textsc{TALYS} default' in Fig.~\ref{fig:results}. The absolute cross-section values are systematically underestimated by a factor of 1.25, which results in $\chi^2_{\mathrm{red}}~=~11.70$. But on average, the energy dependence using this proton+nucleus OMP is better than the one of the \textsc{SMARAGD} calculation. Scaling the calculated cross-section values by 1.25 yields an improved $\chi^2_{\mathrm{red}}$ value of $\chi^2_{\mathrm{red}}~=~3.06$.

Since the `\textsc{TALYS} Bauge $\times$ 2.1' model gives a very good agreement with the experimental data, this model was used to calculate the proton-capture reactivities on $^{130}$Ba. The stellar reactivities $N_A \left<\sigma v\right>^*$ as a function of plasma temperature are given in Table~\ref{tab:reactionrate}. However, the present measurement only covers the high-energy tail of the reaction rate integrand. Due to the rather high nuclear level density of $^{130}$Ba, the ground-state contribution $X$ to the reaction rate is only  $X~=~0.34$ at a temperature of \unit[2.5]{GK} \cite{Rauscher12}. Thus, only \unit[34]{\%} of the reactions will proceed via the ground state, and reactions on excited states dominate the reaction rate. When calculating the stellar reaction rate, one has to assume that the contributions of the thermally excited states to the reaction rate integral are predicted correctly. From this point of view, experimental data at lower energies would be desirable. Nevertheless, the excellent agreement strongly supports the stellar reaction rates as given in Table~\ref{tab:reactionrate}. Stellar reaction rates from the \textsc{NON-SMOKER} code \cite{Nonsmoker} are frequently used in comparison with experimental data, see, $e.g.$ Ref.~\cite{Spyrou13} and for reaction network calculations, see, $e.g.$, Refs.~\cite{Farouqi09, Travaglio11}. Therefore, the stellar reaction rates presented above were compared to the \textsc{NON-SMOKER} results. At the astrophysically relevant temperature of \unit[2.5]{GK}, the newly calculated stellar reaction rate is increased by \unit[68]{\%}.

In order to extent the systematic investigation to test the nuclear-physics input parameters for statistical model calculations, more experimental data in this mass region would be highly desirable. These could include, $e.g.$, proton-induces reactions on the $p$ nucleus $^{132}$Ba or the investigation of proton-capture reactions on the lighter cerium isotopes $^{136,138}$Ce.

\begin{table}[tb]
\caption{Stellar reactivities $N_A \left<\sigma v\right>^*$ for the $^{130}$Ba(p,$\gamma$) reaction as a function plasma temperature.}
\label{tab:reactionrate}
\begin{ruledtabular}
\begin{tabular}{ccc|cc}
%T [GK] & Reactivity [cm$^3$ s$^{-1}$ mole$^{-1}$]\\ \colrule
T & Reactivity && T & Reactivity \\
$\left[\mathrm{GK}\right]$ & [cm$^3$ s$^{-1}$ mole$^{-1}$] && $\left[\mathrm{GK}\right]$ & [cm$^3$ s$^{-1}$ mole$^{-1}$]\\ \colrule
&&& \\
0.15	&	3.529$\times 10^{-35}$	&&2.00	&	4.260$\times 10^{-2}$\\
0.20	&	5.782$\times 10^{-28}$	&&2.50	&	9.965$\times 10^{-1}$\\
0.25	&	1.338$\times 10^{-23}$ 	&&3.00	&	9.805$\times 10^0$\\
0.30	&	1.248$\times 10^{-20}$	&&3.50	&	5.934$\times 10^1$\\
0.40	&	1.035$\times 10^{-16}$	&&4.00	&	2.461$\times 10^2$\\
0.50	&	4.310$\times 10^{-14}$	&&5.00	&	1.756$\times 10^3$\\
0.60	&	3.953$\times 10^{-12}$	&&6.00	&	4.613$\times 10^3$\\
0.70	&	1.428$\times 10^{-10}$	&&7.00	&	5.401$\times 10^3$\\
0.80	&	2.724$\times 10^{-9}$	&&8.00	&	4.014$\times 10^3$\\
0.90	&	3.269$\times 10^{-8}$	&&9.00	&	2.537$\times 10^3$	\\
1.00	&	2.761$\times 10^{-7}$	&&10.00	&	1.562$\times 10^3$	\\
1.50	&	4.621$\times 10^{-4}$	&& &\\

\end{tabular}
\end{ruledtabular}
\end{table}

\section{Summary and conclusion}
Total cross-section values of the $^{130}$Ba(p,$\gamma$)$^{131}$La reaction have been measured at center-of-mass energies between \unit[$E_{\mathrm{c.m.}}~=~3.57$]{MeV} and \unit[$E_{\mathrm{c.m.}}~=~4.96$]{MeV} by means of the activation method. Two clover-type HPGe detectors were used to measure the induced activity. Owing to the high efficiency of the counting setup, it was possible to determine cross-section values for six $\gamma$-ray transitions following the electron-capture decay of $^{131}$La. A reasonable agreement was found between the experimentally determined cross-section values and \textsc{SMARAGD} calculations, although the energy dependence is predicted to be too steep at the lowest energies. The experimental cross-section values are best described by using the proton+nucleus OMP of Ref.~\cite{Bauge01} scaled by a factor of 2.1. This model was then used to give an experimentally supported recommendation for the stellar proton-capture reactivity on the $p$ nucleus $^{130}$Ba. The stellar reaction rate is increased by \unit[68]{\%} compared to the widely used \textsc{NON-SMOKER} results. This measurement extends the scarce experimental data base for charged-particle induced reactions on neutron-deficient nuclei. This can prospectively help to obtain a more globally applicable proton+nucleus OMP.

\begin{acknowledgments}
The authors gratefully acknowledge A. Dewald and the accelerator staff at the Institute for Nuclear Physics in Cologne for providing excellent beams. This project was supported by the Deutsche Forschungsgemeinschaft under contracts DFG (INST 216/544-1, SO907/2-1), the emerging group ULDETIS within the UoC Excellence Initiative institutional strategy, OTKA (PD104664, K108459), and by the HIC for FAIR within the framework of LOEWE launched by the state of Hesse, Germany. GK acknowledges support from the Janos Bolyai Research Scholarship of the Hungarian Academy of Sciences. TR acknowledges support by the Swiss NSF, the European Research Council, and the THEXO collaboration within the 7$^{\mathrm{th}}$ Framework Programme of the EU.
\end{acknowledgments}

\end{document}